\documentclass[useAMS,usenatbib]{mn2e}

\usepackage{graphicx}
\def \mdot   {{\hbox{$\skew3\dot M$}}}

\title[Variable central star in He~2-138]{The dense and asymmetric
central star wind of the young PN He~2-138}
\author[R.K. Prinja et al.]{R. K. Prinja$^{1}$\thanks{E-mail: 
rkp@star.ucl.ac.uk (RKP); seh@star.ucl.ac.uk (SEH); 
urbaneja@ifa.hawaii.edu (MAU),
derck.l.massa@nasa.gov (DLM)},
S. E. Hodges$^{1}$,
M.A. Urbaneja$^{2}$ and D. L. Massa$^{3}$\\
$^{1}$Dept. of Physics {\&} Astronomy, University College London, Gower Street, London WC1E 6BT \\
$^{2}$Institute for Astronomy, University of Hawaii, 2680 Woodlawn Drive, Honolulu, HI 96822, USA \\
$^{3}$Space Telescope Science Institute, 3700 San Martin Drive, Baltimore, MD 21218, USA \\
}
\begin{document}

\date{Accepted 2009. Received 2009; in original form 2009}

\pagerange{\pageref{firstpage}--\pageref{lastpage}} \pubyear{2007}

\maketitle

\label{firstpage}

\begin{abstract}
We present optical {\it ESO} time-series and UV archival
({\it FUSE}, {\it HST}, {\it IUE}) spectroscopy of the
H-rich central star of He 2-138. Our study targets the
central star wind in a very young planetary nebula, and
explores physical conditions that may provide clues to the 
nature of the
preceding post-AGB super-wind phases of the star.
We provide evidence for a dense, slowly accelerating
outflow that is variable on time-scales of hours.
Line-synthesis modelling (SEI and CMFGEN) of low and high ionization UV 
and optical lines is interpreted in terms of
an asymmetric, two-component outflow,
where high-speed high-ionization gas forms mostly in the polar 
region. Slower, low ionization material is then confined primarily to 
a cooler equatorial component of the outflow.
A dichotomy is also evident at photospheric levels.
We also document temporal changes in the weak photospheric
lines of He 2-138, with tentative evidence for a 0.36-day
modulation in blue-to-red migrating features in the absorption
lines.
These structures may betray 'wave-leakage' of prograde non-radial 
pulsations of the central star. These multi-waveband results
on the aspherical outflow of He 2-138
are discussed in the context of current interest in understanding
the origin of axi- and point-symmetric planetary nebulae. 
\end{abstract}


\section{Introduction}
There are several motivations for studying the time-variable
characteristics of PN central stars (CSPN): Chandra and
XMM-Newton observations are providing new perspectives on the
presence of hot bubbles in planetary nebulae formed from the
interaction between central star fast winds and the ambient
asymptotic giant branch (AGB) phase gas (e.g. Kastner et al. 2008).
The origin of the X-ray emission remains uncertain, as does the
consequence of a potentially spatially clumped central star
fast wind. We have previously reported on far-UV
{\it FUSE} satellite spectroscopic time-series observations of
NGC~6543 (Prinja et al. 2007); there we demonstrated in particular
the incidence of coherent, systematically evolving structures
in the fast wind of the H-rich central star. The empirical 
characteristics derived by Prinja et al. (2007) are clearly comparable
to the signatures of large-scale structure in clumped
radiation pressure-driven winds of OB stars.
The post-AGB mass-loss via a fast wind is important not only
because of its dynamical interaction with the nebula, but also
due to its effect on the evolution of the central star.

An improved understanding of the variable nature of CSPN spectral
line profiles is also important in studying binarity in these
objects, and thus post-common envelope evolution. In this case recorded
photometric and spectroscopic signatures may be complicated by
the influence of, for example, variable photospheric structure
and an inhomogeneous inner stellar wind region (see e.g.
De Marco et al. 2008). Furthermore, complex changes in diagnostic
spectral lines can impact on non-LTE model atmosphere analyses aimed
at deriving fundamental parameters of CSPN, including revisions
to mass-loss rates due to strongly clumped outflows
(e.g. Kudritzki, Urbaneja {\&} Puls 2006;
Urbaneja et al. 2008).

Though the UV resonance lines are undoubtedly the most appropriate
for probing CSPN fast winds (e.g. Perinotto 1987), there
is unfortunately a dearth of suitable UV {\it time-series} datasets
({\it IUE}, {\it FUSE}, or {\it HST}) currently available for these
objects. We explore in this paper therefore the alternative of
exploiting optical time-series data that offer diagnostics of
variable conditions in, both, the fast wind and close to
the stellar surface. We focus in this respect on high-quality {\it ESO} 
optical data of the young H-rich central star He~2$-$138.

\subsection{He~2-138}
This paper reports on optical time-series and archival UV data
of the central star of He~2-138 (PK~320-09 1). The central star provides
an interesting balance in that (i) its fast wind reveals
intermediate strength, unsaturated UV resonance lines,
(ii) the weaker optical photospheric lines
are very symmetric and likely not 'contaminated' by wind effects,
and 
(iii)  the wind densities
at low velocities are high enough to provide some optical 
signatures of the outflow.
The nebula itself has a complex knotted appearance, with bubble-like
structures arranged in fairly symmetrical form, providing an overall
elliptical morphology (see e.g. the HST WFPC2 H$\alpha$ imaging
survey of Sahai {\&} Trauger 1998).

There is no recent published model atmosphere
analysis of the central star, though a non-LTE study deriving
a few basic parameters is given by Mendez et al. (1988);
Adopted parameters are listed in Table~1 and include values derived
from a new non-LTE analyses carried out during our present study
of He 2-138.

%%%%%%% Table 1

\begin{table}
 \centering
\caption{He~2-138 central star parameters.}
  \begin{tabular}{lll}
  \hline
Parameter & Value & Reference  \\
\hline

Sp. type & Of (H-rich) & Mendez et al. (1998) \\
T$_{\rm eff}$ & 29000 $\pm$ 1000 K  & This study. \\
Log $g$ & 2.95 $\pm$ 0.15  & This study \\
Log (L/L$_\odot$) & 3.90 & This study \\
Distance & 3.5 kpc & Zhang (1995) \\
Radial velocity & $-$49 km s$^{-1}$ & Schneider et al. (1983) \\
\hline

\end{tabular}
\end{table}

%%%%%%%

Hutton {\&} Mendez (1993) have documented visual magnitude changes
in He 2-138 with amplitudes of between $\sim$ 0.1 and 0.15 mag.
The fluctuations occur over timescales of a few hours and
are similar to those evident in IC~418 (e.g. Mendez et al. 1986).
The magnitude changes are not obviously modulated or periodic and
may relate to variable conditions in the near-photospheric
(inner-wind) regions. In their search for radial velocity variations
in Southern Hemisphere CSPN, Afar {\&} Bond (2005) quote a
greater than 99{\%} probability that He 2-138 is a radial
velocity variable. These changes occur over $\sim$ daily time-scales,
but without any convincing evidence for a period.

Our aim in this study is to determine
optical and UV mass-loss parameters
for an object that is in the early stages of the
planetary nebula phase.
We explore here the evidence for a structured
outflow in He 2-138 and examine corresponding
or unrelated signatures for photospheric variability.

\section{ESO time-series optical spectroscopy}
The study of variability in
the central star of He 2-138 is based primarily on
data collected from the ESO 3.6m telescope at La Silla,
with the HARPS spectrograph, over three nights between
2006 March 24 and March 26 (BJD~2453818.67601 to 2453820.90567).
During each of the three nights 6 spectra were secured over
a time span of $\sim$ 5.5 hours.
The HARPS instrument delivered
a total spectrum range of $\sim$4000 to 7000{\AA}, with
spectral resolution $R$ $\sim$ 110,000. The continuum signal-to-noise
in individual spectra, achieved in 30 minute exposures, is typically 
$\sim$ 40. Standard ESO pipe-line data reduction was conducted using 
MIDAS procedures, including flat-fields, bias subtraction and
wavelength calibration. All spectral line profiles presented here have
been corrected for a system radial velocity of $-$49 km s$^{-1}$
(e.g. Table~1).

%%% Figure 1
\begin{figure}
 \includegraphics[scale=0.37]{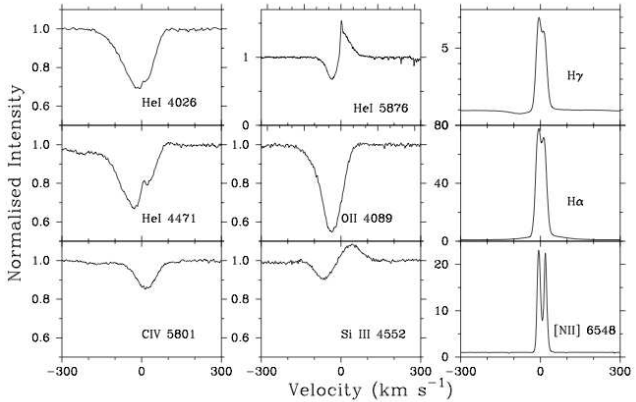}
 \caption{Mean ESO optical spectral lines of photospheric (left-hand
panels),
central star wind (middle panels) and nebular signatures 
(right-hand panels) in He 2-138.}
\end{figure}

The typical spectrum includes stellar lines plus nebular emission.
Characteristic features spanning near-photospheric He~{\sc i}
and weak metal line absorption profiles; blue-shifted absorption
and weak P~Cygni lines; and nebular lines are show in Fig. 1.
The stellar continuum is derived by fitting line-free regions
of the spectrum with a first-order polynomial; this study is
focused on the properties of normalised line profiles and
we do not attempt to derive any flux calibrated measurements.
Though the spectra have been secured in varying seeing conditions
and position angles, the primary line profile changes studied
here can be attributed to stellar variations.  
The nebular lines (e.g. [NII] in Fig. 1) confirm that He 2-138 is a low
excitation planetary nebula.

%The mean full-width at half-maximum
%of the C{\sc iv} $\lambda$5801 and He{\sc i} $\lambda$4026 lines
%are $\sim$ 86 km s$^{-1}$ and 115 km s$^{-1}$, respectively.

\section{Time-averaged fast-wind characteristics}
We collate in this section the overall {\it mean} outflow properties of
He 2-138, based on our ESO optical dataset and archival UV spectra.
Additional fundamental parameters are also derived here from
non-LTE line-synthesis modelling of the time-averaged spectra.

The fast wind of He 2-138 is most sensitively diagnosed in the optical 
time-series primarily by a well developed P~Cygni-like profile in
He{\sc i} $\lambda$5875.57 (2$^3$P$^0$$-$3$^3$D). The blueward
absorption trough extends to $\sim$ $-$200 km s$^{-1}$, with
comparable redward extent for the emission wing. This
line serves as the most useful probe of temporal behaviour in the
fast wind and its variability signatures are derived in Sect.~4.

Additional information on the nature of the fast wind
in He 2-138 can be extracted from UV data. A selection of the
wind-formed line profiles in {\it FUSE}, {\it IUE}, and {\it HST}
spectra is shown in Fig. 2. The wind is evident in low-ionisation
species due to C{\sc ii} $\lambda$1335,
Al{\sc iii} $\lambda\lambda$1855,1863,
Mg{\sc ii} $\lambda\lambda$2796, 2802, and also in the higher ions
S{\sc iv} $\lambda\lambda$1063, 1073,
Si{\sc iv} $\lambda\lambda$1394, 1403, and
C{\sc iv} $\lambda\lambda$1548, 1551. Note that
N{\sc v} $\lambda\lambda$1238, 1242 and
O{\sc vi} $\lambda\lambda$1032, 1038, which generally trace shocked gas
in hot star winds, is absent in He 2-138. The blueward absorption
trough of the high-ion lines extend to $\sim$ 500 to 700 km s$^{-1}$,
and $\sim$ 100 to 300 km s$^{-1}$ in the low ion species.
Overall, the wind lines exhibit strong optical depths, including
saturated regions at low to intermediate velocities.

%%% Figure 2
\begin{figure}
 \includegraphics[scale=0.5]{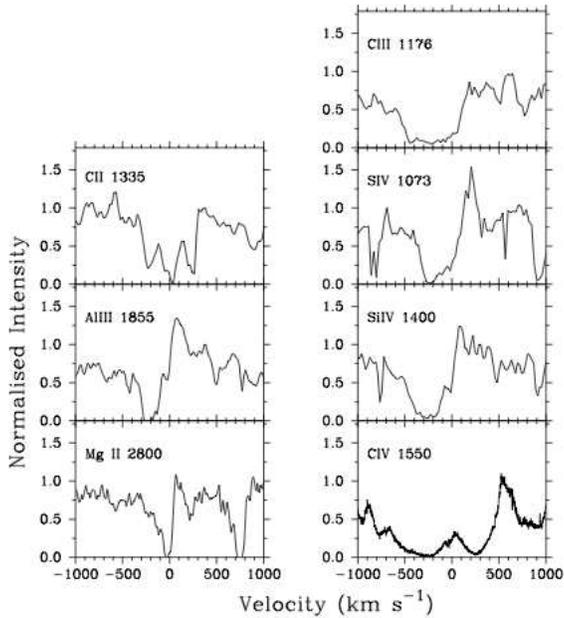}
 \caption{{\it FUSE}, {\it IUE} and {\it HST} spectral lines that provide
UV signatures of the fast-wind in He 2-138.}
\end{figure}

%%%%%%%

The impression from the UV lines and a P~Cygni-like He{\sc i} 
$\lambda$5876
is that He 2-138 has a dense, comparatively slow central star wind.
Further support is provided by the unusual behaviour of the
Si{\sc iii} $\lambda$1300 triplets shown in Fig. 3,
where the rest wavelengths of the lines are indicated in
the radial velocity corrected {\it IUE} spectrum.
The Si{\sc iii} triplets are clearly blueshifted by between
$-$60 km s$^{-1}$ to $-$80 km s$^{-1}$ and thus trace the
radial expansion at the base of the outflow. We conclude
that a very dense, slowly accelerating wind is implied in these data.

%%% Figure 3
\begin{figure}
 \includegraphics[scale=0.39]{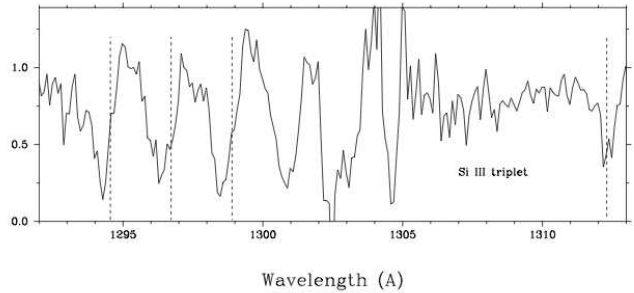}
 \caption{Blueward shifted Si{\sc iii} $\lambda$1300 triplets in the
UV spectrum of He 2~138. The Si{\sc iii} singlet at $\lambda$1312
is also shown for comparison. The vertical dashed lines mark
the rest positions.}
\end{figure}

%%%%%%%

%%% Figure 4
\begin{figure*}
 \includegraphics[scale=0.3]{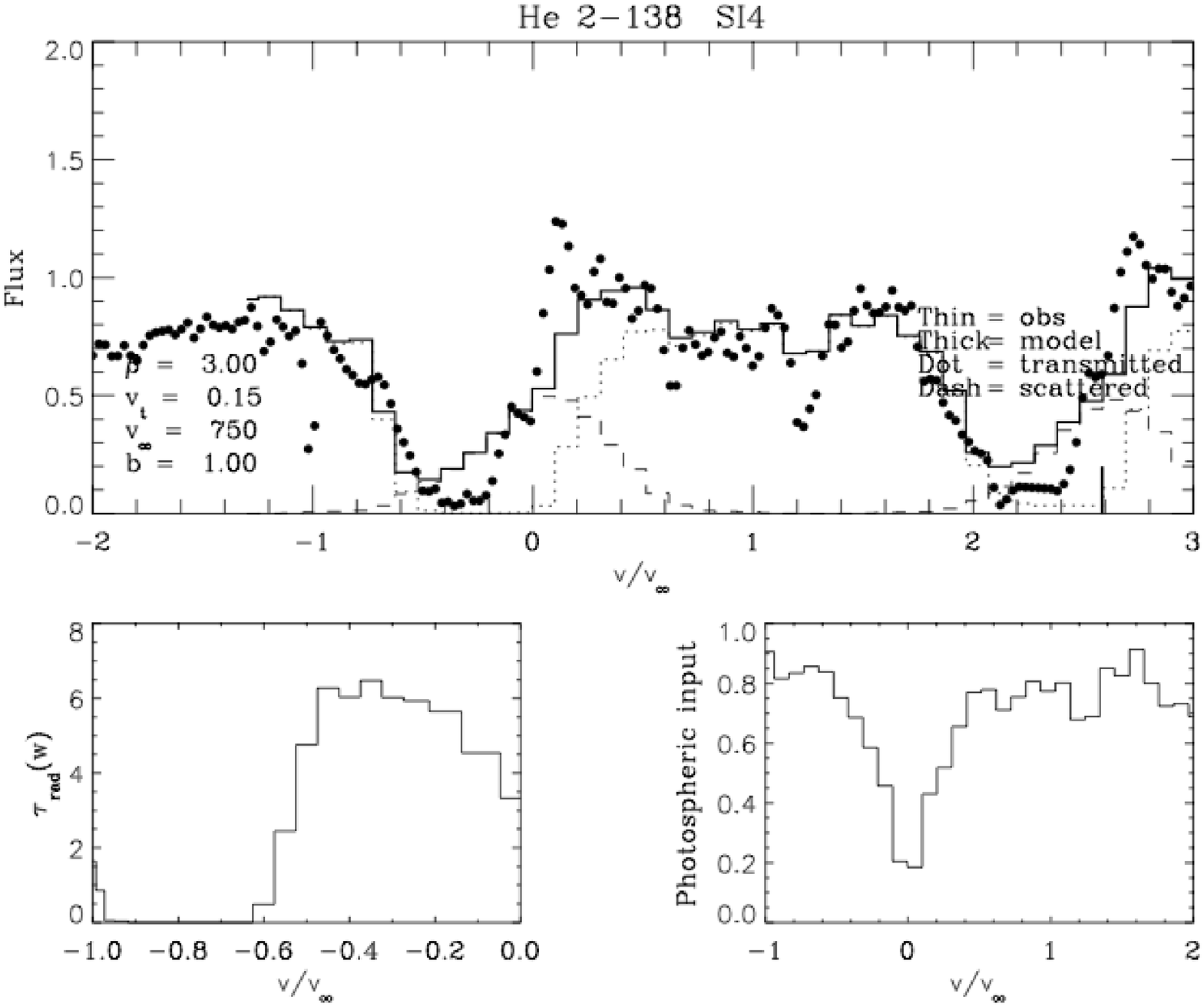}
 \includegraphics[scale=0.3]{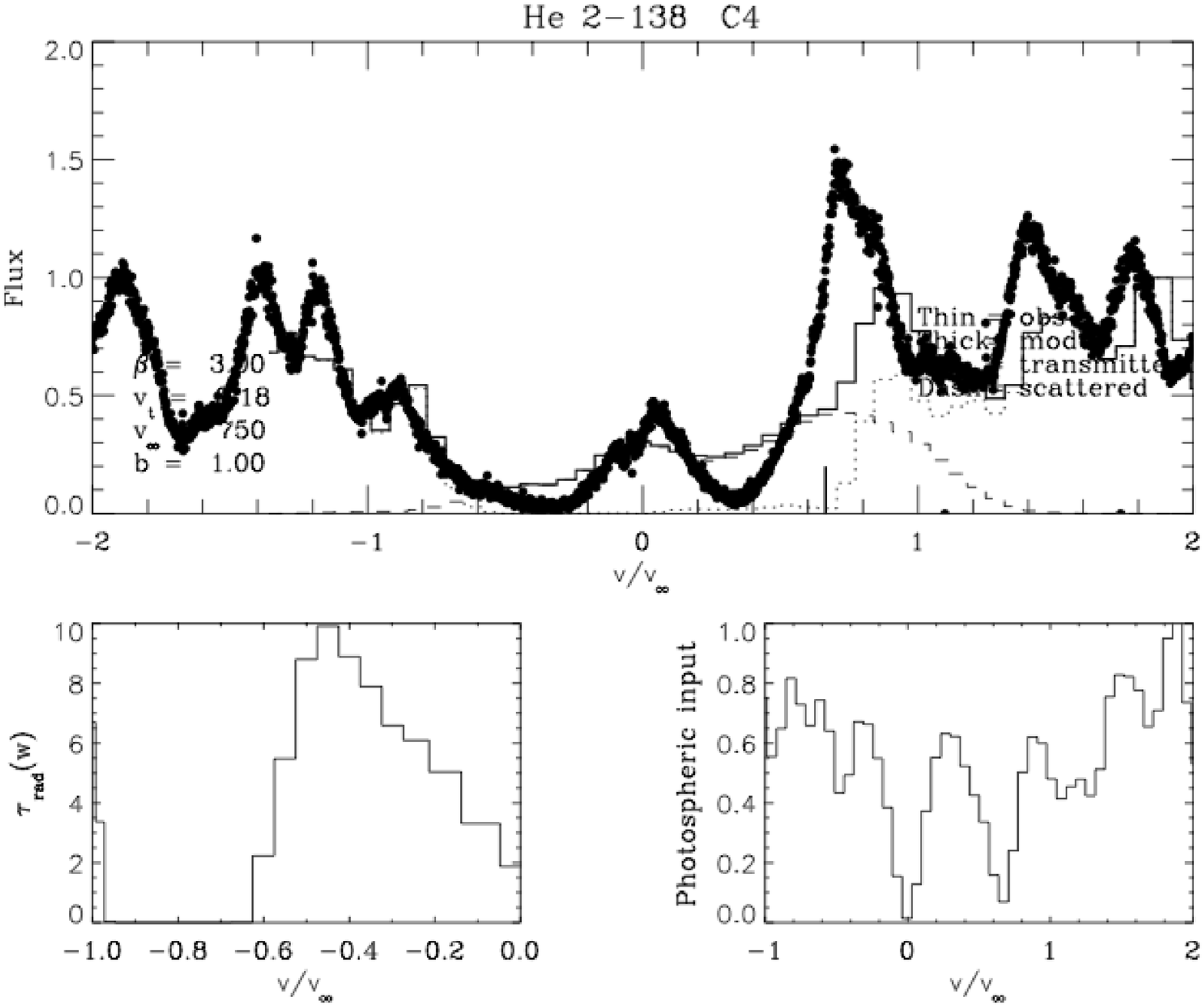}
 \includegraphics[scale=0.3]{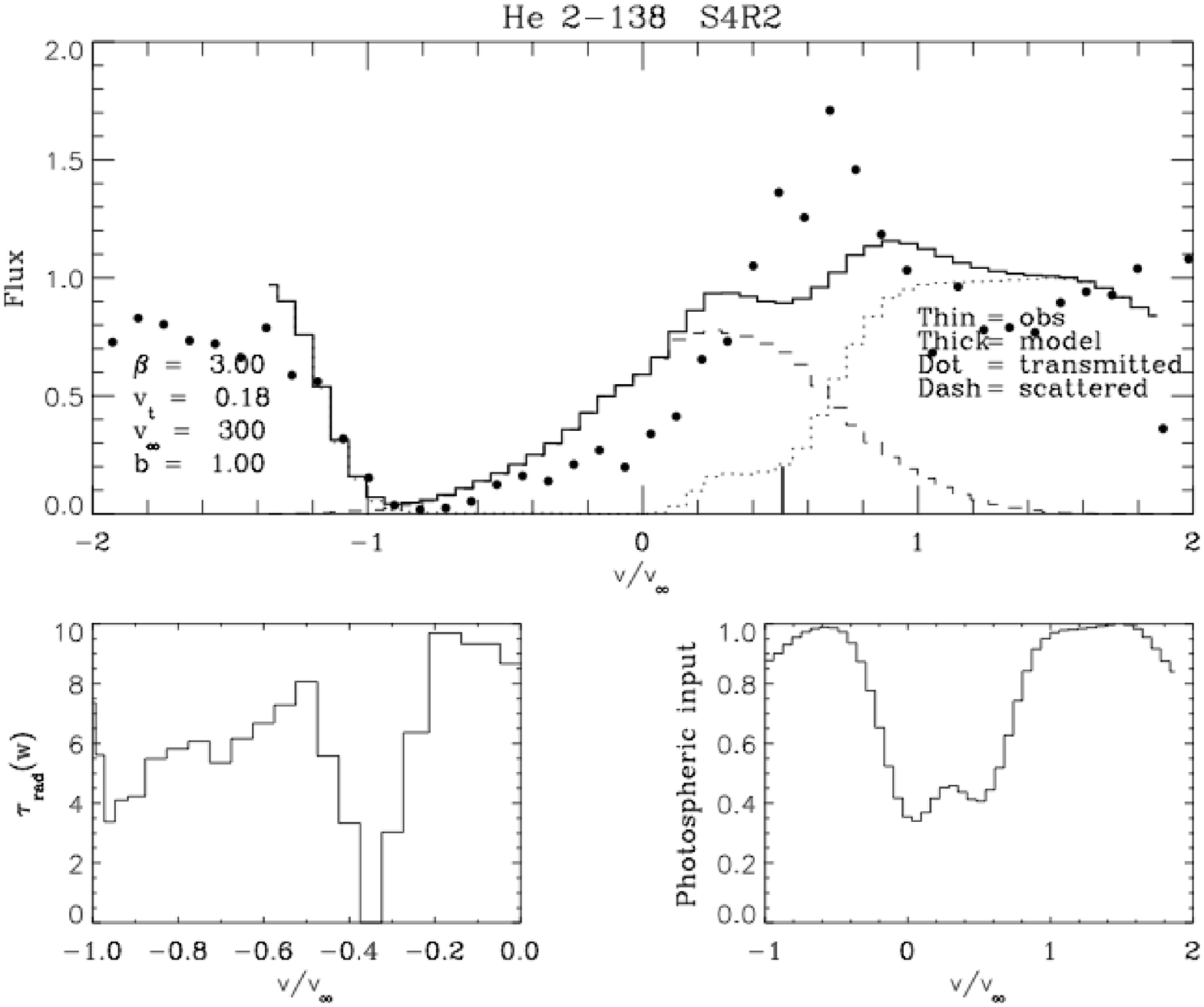}
 \includegraphics[scale=0.26]{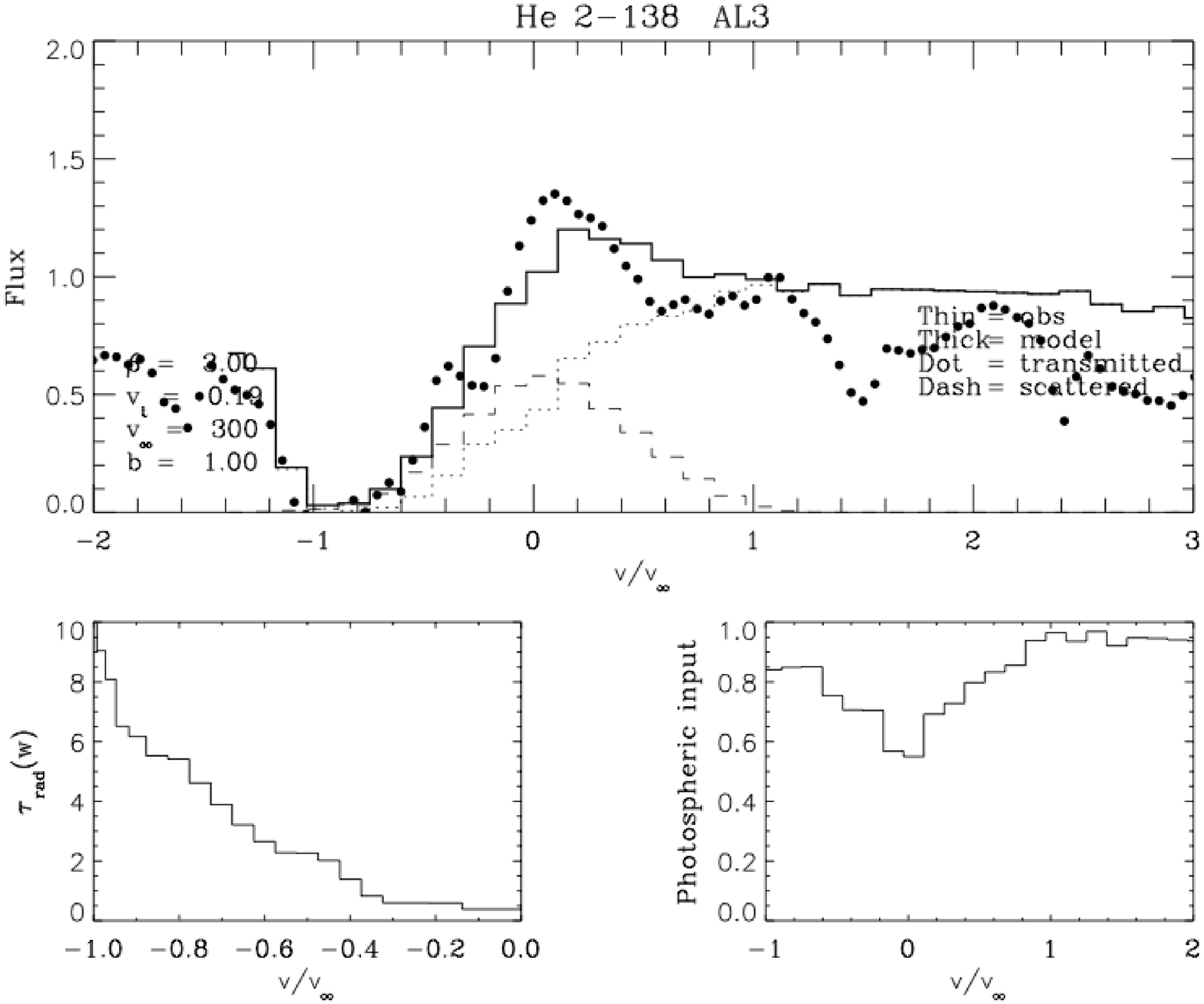} 
\caption{SEI model fits (solid line) to Si{\sc iv} $\lambda\lambda$1400 
({\it IUE}),C{\sc iv} $\lambda\lambda$1550 ({\it HST}),
S{\sc iv} $\lambda$1073, and Al{\sc iii} $\lambda$1855
({\it FUSE}) P Cygni profiles
of He 2-138. The panels below each fit show the adopted radial 
optical depths and input (TLUSTY) photospheric spectra.
Note that the maximum velocity adopted in S{\sc iv} 
and Al{\sc iii} of
300 km s$^{-1}$ has to be raised to 750 km s$^{-1}$ to match
the blue wings of Si{\sc iv} and C{\sc iv}}.
\end{figure*}

%%%%%%%

%%% Figure 5
\begin{figure*}
 \includegraphics[scale=0.65]{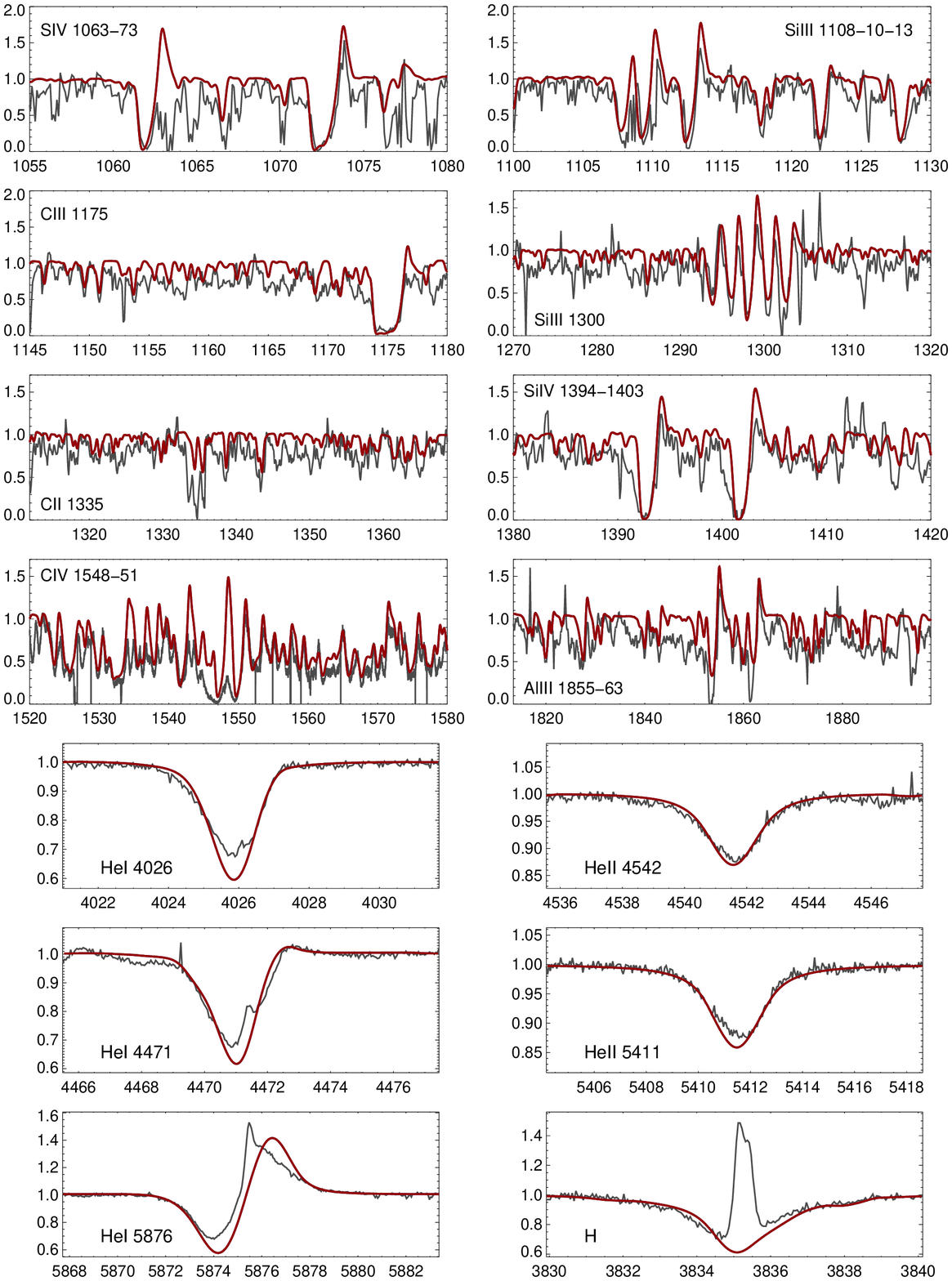}
 \caption{CMFGEN non-LTE multi-line fit (red line; Table 3) to strategic
UV and optical spectral feature of He 2-138. This low terminal
velocity case ($v_\infty$ = 300 km s$^{-1}$) provides
a suitable match to the FUV and low ionization lines, but
the model cannot adequately reproduce the strong Si{\sc iv}
and C{\sc iv} resonance lines.}
\end{figure*}

%%%%%%%

\subsection{Line-synthesis modelling}
To decode physical parameters from the UV wind resonance lines
we first employed the 'Sobolev with exact integration' (SEI) code
of Lamers et al. (1987), but with modifications detailed
in Massa et al. (2003); these changes principally permit
the radial optical depths of the wind to be treated as
independent, discrete and variable bins in velocity space.
Once the empirical velocity law of the wind
is determined, this approach provides reliable radial optical
depths for unsaturated wind
lines as a function of normalized velocity, $\tau_{rad}(w)$, where
$w = v/v_\infty$ and $v_\infty$ is the terminal velocity of the wind.  
The
fits to individual wind lines require parameters for the terminal
velocity, `$\beta$-type' velocity law, turbulent velocity, and
$\tau_{rad}(w)$ (specified as a set of 21 independent velocity bins).
The effect of a photospheric spectrum in the SEI profile fitting is
approximated by adopting a TLUSTY plane-parallel model atmosphere 
spectrum (Hubeny {\&} Lanz, 1995) for
T$_{\rm eff}$ = 29000 and log$g$ = 3.0
and a projected rotation velocity of 45 km s$^{-1}$ (see below and
Table~1).
This input photospheric spectrum provides a reasonable
match to the 'underlying' absorption and it primarily
needs to be incorporated to match wind-formed profiles that
are located across the heavily line-blanketed UV regions
between $\sim$ 1300{\AA} to 1900{\AA}.
The $\tau_{rad}(w)$ values were
converted into a product of $\mdot$ times ionization fraction,
$\mdot\,q_i(w)$ (see, e.g., Massa et al.\ 2003); we
derive here the mean $\langle\mdot\,q_i\rangle$ values over
0.2 $\le$ $v/v_\infty$ $\le$ 0.9 for several ions analysed.

SEI model fits to Si{\sc iv} $\lambda\lambda$1400 ({\it IUE}),
C{\sc iv} $\lambda\lambda$1400 ({\it HST}),
S{\sc iv} $\lambda$1073 ({\it FUSE}),
and Al{\sc iii} $\lambda\lambda$1855, 1863 ({\it IUE})
are shown in Fig. 4,
and include panels displaying the TLUSTY photospheric inputs
and the derived radial optical depths as function of
normalised velocity. We note firstly that reasonable matches
to the observed data are obtained, but we require a different
maximum velocity for Si{\sc iv} and C{\sc iv} (= 750 km s$^{-1}$)
as opposed to 300 km s$^{-1}$ for S{\sc iv} and the low
ion species generally.
For an assumed solar abundance and parameters adopted in Table~1
(and derived below),
the values obtained for the product of mass-loss
rate and ionization fraction ($\langle\mdot\,q_i\rangle$)
are listed in Table 2.

%%%%%%% Table 2

\begin{table}
 \centering
\caption{SEI-derived mass-loss and wind ionization parameters.}
  \begin{tabular}{lll}
  \hline
Ion & $\langle\mdot\,qi\rangle$ & $\langle{q_i}\rangle$ \\
 & (M$_\odot$ yr$^{-1}$) & \\
\hline

C$^{3+}$ & 6.9 $\times$ 10$^{-11}$ & 6 $\times$ 10$^{-4}$ \\
Si$^{3+}$ & 1.7 $\times$ 10$^{-10}$ & 1 $\times$ 10$^{-3}$ \\
S$^{3+}$ & 2.4 $\times$ 10$^{-9}$ & 2 $\times$ 10$^{-2}$ \\
Al$^{2+}$ & 5.5 $\times$ 10$^{-10}$ & 5 $\times$ 10$^{-3}$ \\
C$^{+}$ & 1.1 $\times$ 10$^{-11}$ & 9 $\times$ 10$^{-5}$ \\
\hline

\end{tabular}
\end{table}

%%%%%%%

Despite the
slow velocity law used ($\beta$ = 3) and generally high
turbulent velocities ($v_t$), the SEI models mostly under-predict
the strengths of the emission components. Additionally, the
fit to the C{\sc iv} doublet remains problematic, where the redward
low-velocity absorption component is also too weak in the model.
It is also unexpected to note the large differences between the
S{\sc iv} $\lambda\lambda$1063, 1073 and
Si{\sc iv} $\lambda\lambda$1394, 1403 lines, given that the
ionization ranges of these lines bracket each other so closely
(though the Si line has greater cosmic abundance).
These discordances (Fig. 4) may confirm our conclusion
earlier that He 2-138 has a very dense wind. It might in fact be
that the density of the wind is so high (with a very shallow
velocity law) that the Sobolev (single scattering)
theory in no longer valid and the normal SEI source function is
not appropriate in this case.

To gain further insights into the time-averaged characteristics
of the wind, we also examined line synthesis using the
unified non-LTE, line blanketed model atmosphere code CMFGEN (e.g. Hillier 
{\&} Miller 1998).
Very 
briefly, CMFGEN solves the non-LTE radiation transfer problem assuming a 
chemically homogeneous, spherical-symmetric, steady-state outflow. Each 
model is defined by the stellar radius, the luminosity, the 
mass-loss rate, the wind terminal velocity (v$_\infty$), the 
stellar mass, and by the abundances of the species included in the 
calculations. The code does not solve for the hydrodynamical structure, 
hence the velocity field has to be defined. A solution, commonly adopted, 
is to use the output of a plane-parallel model (such as TLUSTY) to define 
the pseudo-static photosphere, connected just below the sonic point to a 
beta-type velocity law to describe the wind regime. We consider H, 
He~{\sc i/ii}, C~{\sc ii/iii/iv}, N~{\sc ii/iii/iv/v}, O~{\sc 
ii/iii/iv/v}, Mg~{\sc ii/iii}, Al~{\sc iii/iv}, Si~{\sc iii/iv}, P~{\sc 
iv/v}, S~{\sc ii/iv/v/vi} and Fe~{\sc iii/iv/v/vi/vii} in our models.

The ionization balance of He is used to derive the effective temperature. 
We consider only He{\sc i} lines belonging to the triplet spin system and 
He{\sc ii} lines present in the optical spectrum. In this case, the 
He{\sc ii} lines react strongly to changes of $\pm$\,1000 K, whilst the 
He{\sc i} lines remain unaffected. The surface gravity is determined by 
fitting the H Balmer lines series. We give more weight to higher lines in 
the series, because the effect of the stellar wind and the contamination 
by the ionized gas is less important. It has to be noted that the 
photospheric structure is linked to the stellar mass, hence its adopted 
value could produce changes. These would 
be very small, however, for our target object.

%%%%%%% Table 3

\begin{table}
 \centering
\caption{CMFGEN best solution parameters.}
  \begin{tabular}{lll}
  \hline
Parameter & Value \\
\hline

T$_{\rm eff}$ & 29000 $\pm$ 1000 K  \\
Log $g$ & 2.95 $\pm$ 0.15\\
Log (L/L$_\odot$) & 3.90 \\
N(He)/N(H) & 0.20 \\
log Q & $-$11.45 \\
$\mdot$ & 1.23 $\times$ 10$^{-7}$ M$\odot$ yr$^{-1}$ \\
Chem. composition & solar \\
\hline

\end{tabular}
\end{table}

%%%%%%%

As shown in Fig. 5, our final model (Table 3) reproduces well most of the 
features present in the FUV/UV spectrum. However, the model does not 
match the wide absorption troughs observed in the Si~{\sc iv} and C~{\sc 
iv} UV 
profiles, and it under-predicts the
strength of the Al~{\sc iii} lines. Low ionization 
species present in the UV range, such as C~{\sc ii} and Al~{\sc iii} will 
require a lower (significantly lower in the case of the C features) 
effective temperature, but will then not reproduce the observed 
He~{\sc i/ii} 
ionization balance. Similarly, high ionization features present in the 
optical spectrum (O~{\sc iii} and C~{\sc iv} lines) would require a 
higher effective temperature, that again would not be consistent with the 
He ionization balance.
We noted the same discrepancy in the SEI modelling earlier
(Fig. 4), where a 'two component' wind was invoked and
$v_{\infty}$ = 750 km s$^{-1}$ used for the C~{\sc iv} 
and Si~{\sc iv} {\it IUE} lines.
Note that the {\it HST} STIS spectrum of C{\sc iv} modelled in Sect. 3.1
(Fig. 4) was secured in 2004 June and is broadly similar to
the {\it IUE}
high-resolution line profile taken in 1989 May. The maximum
epoch to epoch change
evident in the archival data is $\sim$ 100 km s$^{-1}$ in the
shallow blueward wing. The variation is thus not sufficient to
account for the two component behaviour discussed above.

Overall, the unusual nature of the wind lines in He 2-138
is borne out by our comparisons to the model line profiles
for spherically symmetric winds. We believe that the
dichotomy in matching low and high ion lines,
including discrepancies (with respect to
symmetric models) in the
relative absorption and emission strengths of the P Cygni profiles,
argue in favour of a wind whose terminal velocity and mass-loss have
a latitude dependence.
Low velocity, low ion species would form primarily in
the cooler equatorial regions of an asymmetric outflow that is
viewed almost pole-on. The high speed Si{\sc iv} and
C{\sc iv} bearing gas is then driven out mostly in the hotter polar
regions of the flow. This notion of a two-component, asymmetric
wind in He 2-138 is discussed further in Sect. 5

For a $v_{\infty}$ of 300 km s$^{-1}$, $T_{\rm eff}$ = 29000 K,
adopted
central star mass of $\sim$ 0.6 M$_\odot$ and
a log (L/L$_\odot$) $\sim$ 3.9,
the model in Fig. 5 yields a mass-loss (wind strength) parameter
log Q $\sim$ $-$11.45 dex
(where Q = $\mdot$(R$_{\star}$$v_\infty$)$^{-1.5}$)
and thus a mass-loss rate of $\sim$ 1.2 $\times$ 10$^{-7}$
M$_\odot$ yr$^{-1}$. The adopted model is for a smooth wind
and any clumping is unconstrained here, but it is noted as being at a
very low degree.
The corresponding wind momentum value
log ($\mdot\,v_{\infty}\,R_{\star}^{0.5}$) of $\sim$ 26.6 dex sits
within the scatter in the wind-momentum luminosity relation observed
for CSPN and predicted by the line-driven wind theory
(see e.g. Kudritzki, Urbaneja {\&} Puls, 2006).
Though these authors note that in the luminosity regime
restricted to CSPN (and excluding luminous O-type stars), there
is no convincing relation between wind momentum and luminosity.

It is interesting to compare this result from the {\it ab initio}
CMFGEN prediction to the empirical, SEI-derived
$\langle\mdot\,q_i\rangle$ results presented earlier.
For  a mass-loss rate of $\sim$ 1.2 $\times$ 10$^{-7}$ M$_\odot$ 
yr$^{-1}$, the
corresponding
mean ion fractions based on the SEI profile fits (Fig. 4)
are also listed in Table 2.
The empirical analyses would suggest therefore that none
of these strongly absorptive ions are dominant
in the outflow of He 2-138. We note however that the
CMFGEN base model (Table 3) predicts (for smooth or
weakly clumped winds) that the S$^{3+}$ and Si$^{3+}$
ionization fractions are dominant in the wind (i.e.
$\sim$ 90{\%}). This disparity would suggest that
either the CMFGEN derived mass-loss rate is overestimated
or the SEI product of $\mdot\,q_i(w)$ is underestimated.

%%% Figure 6
\begin{figure}
 \includegraphics[scale=0.35]{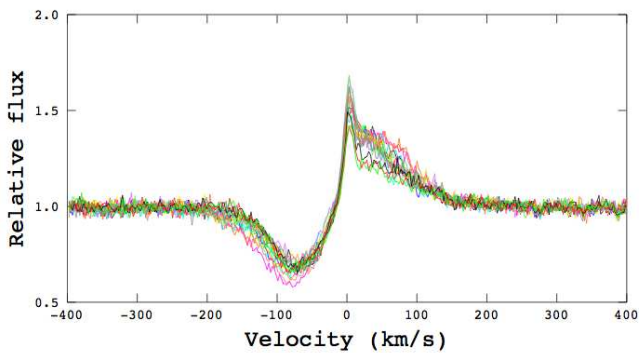}
 \caption{Variability evident in the 18 {\it ESO}
He{\sc i} $\lambda$5876 P Cygni profiles of He 2-138,
over $\sim$ 2.2 days.}
\end{figure}

%%%%%%%

\section[]{Temporal changes in the wind and photosphere}
The time-dependent nature of the central star wind in He~2-138 is
discussed here, based primarily on the ESO time-series data of
the P~Cygni-like He{\sc i} $\lambda$5876 feature (see Fig. 1).
We also comment on additional changes noted in the deep-seated,
photospheric dominated, optical lines due to He{\sc i} and
C{\sc iv} $\lambda$5801. Unfortunately there is no appropriate FUV
or UV time-series dataset currently available for He~2-138.
We examined the fragmented sample of 12 {\it FUSE} (LWRS) spectra
present in the archive, covering epochs in 1999 September,
2000 March, 2000 July, 2002 April, 2007 February and 2007 April.
There is no evidence for substantial blueward variability in
for example the S{\sc iv}, P{\sc v} or C{\sc iii} lines. The
maximum discernible change is a $\sim$ 60 km s$^{-1}$ shift in
the blue wing of S{\sc iv}$\lambda$1073.5.

%%% Figure 7
\begin{figure*}
 \includegraphics[scale=0.55]{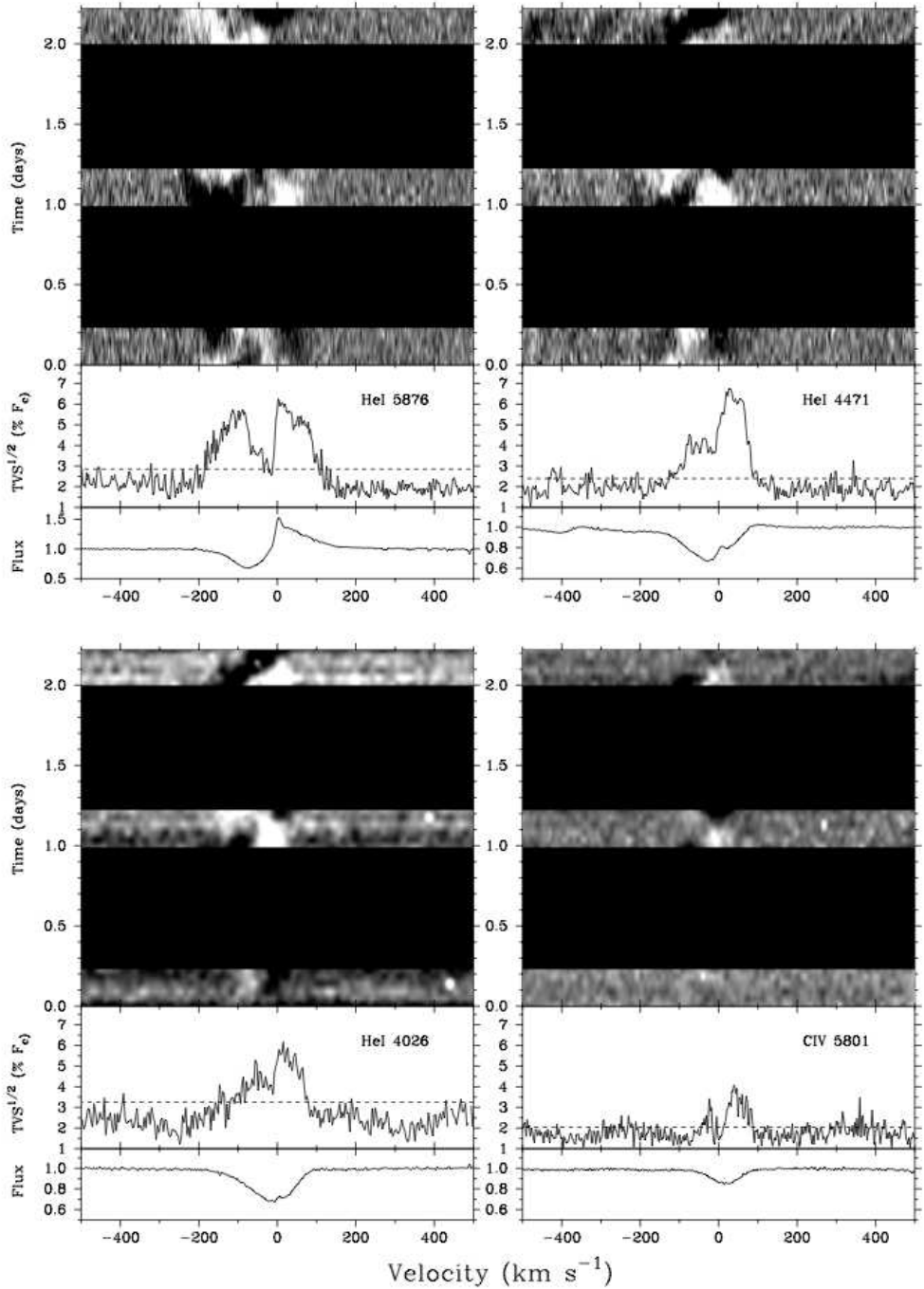}
 \caption{Temporal characteristics of the wind dominated
He {\sc i} $\lambda$5876 and photospheric
He {\sc i} $\lambda$4026}, He {\sc i} $\lambda$4471
and C {\sc iv} $\lambda$5801 lines.
The grayscale images show temporal changes in individual spectra 
normalised by the
mean profile shown at the bottom for each line. The dynamical
range shown is 0.96 (black) to 1.05 (white). The middle panels for
each line provide the variance statistic, where we highlight
changes above the 95{\%} significance level (dotted line).
\end{figure*}

%%%%%%%

Unambiguous evidence for changes in the central star wind
comes from the He{\sc i} $\lambda$5876 line. The sequence of
18 spectra secured over $\sim$ 2.2 days is shown in Fig. 6.
The absorption trough is variable out to $\sim$ 200 km s$^{-1}$.
The equivalent widths vary between $\sim$ 0.37{\AA} to 0.65{\AA}
and $\sim$ 0.47{\AA} to 0.8{\AA}, for the absorption and
emission components, respectively. 
These changes are correlated such that the deepest
absorption troughs are accompanied by the strongest emission
components.
At its greatest, the absorption and emission equivalent widths
of the profile can increase (in concert) by $\sim$ 60{\%}.

The temporal characteristics of He{\sc i} $\lambda$5876 line profile are
explored further in Fig. 7, together with the
properties of the photospheric dominated
absorption lines of He{\sc i} $\lambda$4026, He{\sc i} $\lambda$4471,
and C{\sc iv} $\lambda$5801.
The Temporal Variance Spectrum (TVS; e.g. Fullerton et al. 1996)
indicates a slight blueward asymmetry in the profile changes
of He{\sc i} $\lambda$5876 at the 95{\%} significance level.
A strong absorption enhancement is evident during the second night
of observations, extending to $\sim$ 200 km s$^{-1}$. These data
are not intensive enough to conclude whether this is an
episodic migrating feature akin to the discrete absorption
components (DACs) seen generally in the UV spectra of OB stars
and also reported in the {\it FUSE} data of NGC~6543 by
Prinja et al. (2007). The weaker photospheric lines are
also undoubtedly variable on hourly time-scales and the grayscale
representations in Fig. 7 reveal similar systematic changes
in each of three lines considered.
The equivalent widths of the photospheric lines are not conserved,
and we measure changes of up to $\sim$ 25{\%} in the
He{\sc i} $\lambda$4471 and He{\sc i} $\lambda$4026 spectra.
These variations do not correlate with the blueward or
redward fluctuations of the He{\sc i} $\lambda$5876 P~Cygni profile.
The same is true for
the very weak and narrow Si {\sc iii} $\lambda\lambda$4552, 4567 and 4574
triplet shown in Fig. 1, which exhibits $\sim$ hourly changes that
mimic those of He{\sc i} $\lambda$4471 and He{\sc i} $\lambda$4026.
We conclude that temporal changes in the central star wind
are not evident in the deep-seated regions close to the photosphere
and likely arise further out in the outflow, where they are
diagnosed via He{\sc i} $\lambda$5876.

In Fig. 7, the data of photospheric lines, which are shown as
residuals from the mean profiles,
reveal blue to red travelling pseudo-absorption
and pseudo-emission features in the dynamical spectra.
This behaviour may be indicative of the presence of prograde
non-radial pulsations of the photosphere. The variability
pattern is similar from night to night.
In each of the photospheric lines the redward variance (TVS)
is stronger than the blueward TVS, and this asymmetry is
marked by a drop in variance at rest velocity.
The blue-to-red features
migrate across the full span of the absorption trough
over an estimated $\sim$ 0.3 to 0.4 days.
The empirical changes in the near-surface lines of
e.g. He{\sc i} $\lambda$4471, He{\sc i} $\lambda$4026
and C{\sc iv} $\lambda$5801 are characterised by localised
structure in the absorption troughs as opposed to whole-scale
radial velocity shifts about line centre. These sub-features
can have the effect of extending either the blue or red wing
of the profile, while the other wing stays constant. In extreme
the features can results in a $\sim$ 25{\%} change in
equivalent width. Note that though the variations in
He{\sc i} $\lambda$4471 and He{\sc i} $\lambda$4026 (and
other similar features in the optical spectrum of He 2-138)
extend to more that $\pm$ 100 km s$^{-1}$ (Fig. 7), the basic
pattern of changes is the same as that evident in
C{\sc iv} $\lambda$5801, which is a much narrower line
(closer to $\pm$ projected rotation velocity) and likely
forms very close to or at the photosphere.

With the caveat that
our time-series dataset of He 2-138 is sparse, we conducted
a periodogram analysis to search for evidence of cyclic behaviour
in the He{\sc i} $\lambda$4026 and 
He{\sc i} $\lambda$4471 lines. The frequency range
sampled by the data from $\sim$ 0.5 days$^{-1}$ to
10 days$^{-1}$ was examined for periods by employing the
CLEAN algorithm described by Roberts et al. (1987). Following an
iterative deconvolution of the (strong) window function for our
dataset, the only potentially significant peaks in the resulting power
spectra are aliases of each other at $\sim$ 1.4 d$^{-1}$
and 2.8 d$^{-1}$ (see Fig. 8).
We estimate a $\sim$ 30{\%} uncertainty in these frequencies
based on the widths of the power spectra features.
The shorter corresponding
period, i.e. 0.36 days, is repeated over more than six cycles
during the full span of the observing run. Grayscale representations
of the 18 individual quotient (with respect to the mean) line
profiles are shown phased on the 0.36 day period in Fig. 9
(folded over 2 cycles). The He{\sc i} $\lambda$4026 and
He{\sc i} $\lambda$4471 lines clearly show some coherency in
the travelling features over the 0.36 day time-scale. In contrast
the outflow dominated He{\sc i} $\lambda$5876 profile is
more fragmented in the phased plots and there is certainly no
evidence that the outflow changes are modulated on this time-scale.

The detection of a residual variability pattern in Figs. 7 and 9 that
extends to velocities higher than the mooted projected rotation
velocity may be interpreted in terms of the 'wave-leakage'
mechanism, which permits pulsations to initiate
perturbations in the deepest stellar wind regions.
For example, Cranmer (1996) and Towsend (2000) argue that
pulsations arising in the stellar interior can migrate
toward the outer atmospheric layers as high frequency gas
pressure (p mode) waves or as lower frequency (g mode) waves.
The notion of wave-leakage in He 2-138 would be consistent
with our conclusion in Sect. 3.1 that the central star has a
very dense wind, thus in effect there is no clear boundary
at which pulsations get reflected and confined.

\section[]{Discussion}
We have revealed that He 2-138 has a dense, slowly accelerating
central star wind, that is variable on time-scales of hours to days.
The SEI and CMFGEN line synthesis models presented here
support the notion that He 2-138 exhibits a 'two-component'
outflow: A hotter (i.e $>$ 29,000 K) component provides
a better match to the (near) photosheric
C{\sc iv} and O{\sc iii} optical lines, but
predicts too much He{\sc ii} and hence less He{\sc i}.
In contrast cooler models ($\sim$ 25,000 K)
reproduce better the low-ionization UV lines (including
C{\sc ii} and Al{\sc iii}), but these models predict
too much photospheric He{\sc i} and too weak He{\sc ii} lines.
Additionally, the spherically symmetric SEI and CMFGEN models
cannot  reproduce the observed relative absorption and emission strengths 
of the Si{\sc iv} $\lambda\lambda$1400,
C{\sc iv} $\lambda\lambda$1550,
and Mg{\sc ii} $\lambda\lambda$2800 P Cygni profiles.
This discordance may be interpreted to suggest that the outflow
is asymmetric.
While the UV
S{\sc iv} $\lambda$1073 line is (a lower abundance) surrogate to
Si{\sc iv} in the range of ionization potential, its
line profile can only be matched by adopting a maximum wind velocity
of 300 km s$^{-1}$, while a higher velocity of 750 km s$^{-1}$ is
needed for Si{\sc iv} $\lambda\lambda$1400 (and
C{\sc iv} $\lambda\lambda$1550). Though the spectral lines
we considered have been secured at different epochs (for
{\it ESO}, {\it IUE}, {\it HST}, and {\it FUSE} data), multiple
archival FUV and UV spectra spanning several years do not provide
any indication that the shortward wings of the wind lines are
variable by several 100 km s$^{-1}$.

%%% Figure 8
\begin{figure}
 \includegraphics[scale=0.5]{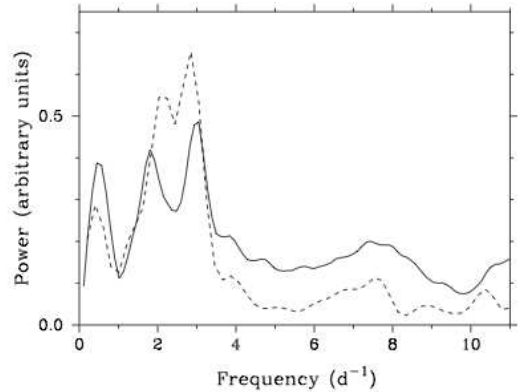}
 \caption{Power spectrum (in arbitrary units) for
He{\sc i} $\lambda$4026 (solid line) and He{\sc i} $\lambda$4471
(dashed line). The strongest peaks correspond to (alias) signals
at $\sim$ 0.36 days and 0.72 days.}
\end{figure}

An analogous example of a 'hybrid' wind may be
that of B[e] stars, where a two-component stellar wind model
has been proposed (e.g. Zickgraf et al. 1985).
In this scenario a hotter and faster polar wind gives rise to
more extended high-excitation absorption features, and in contrast
a slowly accelerating cooler equatorial disk wind promotes
narrower low-excitation lines. Extremely slow (less than 100 km s$^{-1}$)
disk winds can then be seen in B[e] stars viewed edge on.
The increasing maximum outflow velocity in He 2-138 from
Al{\sc iii} to S{\sc iv}, and to Si{\sc iv} and C{\sc iv} may reflect
the spatial distribution of these ions in a hypothesised latitude
dependent outflow viewed approximately pole-on.
One inconsistency in comparison to B[e] star-type hybrid winds
is that the N{\sc v} $\lambda\lambda$1240 doublet would in this
scenario be expected to be strongly in emission. This resonance
line is however absent in {\it IUE} spectra of He 2-138,
and since {\it FUSE} data {\it do} reveal the
presence of N{\sc iii} $\lambda$955, we do not anticipate a strong
N abundance anomaly. The absence of N{\sc v} may instead reflect
a general lack of shocked X-ray emitting
gas. The O{\sc vi} $\lambda\lambda$1032, 1038 doublet is also
not present.
N$^{2+}$ is then expected to be a dominant ionisation stage in the wind.

In addition to a proposed non-spherical outflow in He~2-138,
our analyses also provides indications of some dichotomy
in the photospheric lines (see above). One possibility is that
the central star is a rapid rotator, and we are indeed viewing it
almost pole on (i.e. at a smaller projected velocity).
The rapid rotation distorts the surface layers sufficiently
such that we view a hotter polar region and cooler equatorial
surface.

%%% Figure 9
\begin{figure*}
 \includegraphics[scale=0.6]{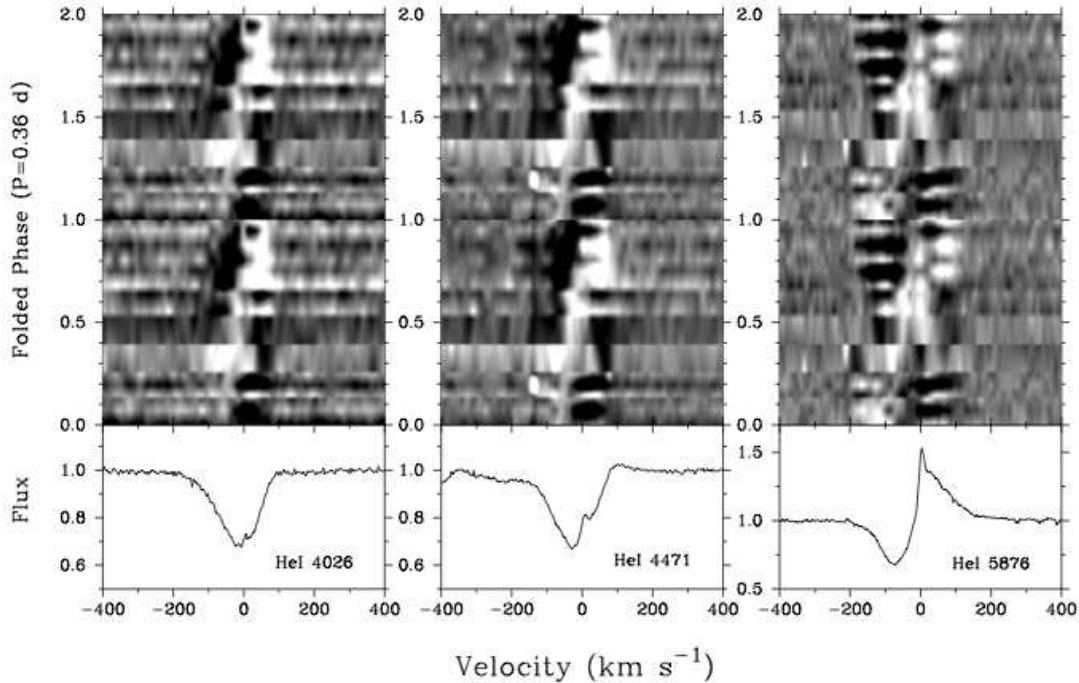}
 \caption{Individual He{\sc i} $\lambda$4026,
He{\sc i} $\lambda$4471} and He{\sc i} $\lambda$5876 quotient
spectra are shown phased over 2 cycles on a period of
0.36 days.
\end{figure*}

%%%%%%%

Evidence for asymmetries in the central star wind of a very young
PN such as He 2-138 is interesting in connection to the wider
debate as to the origin of commonly seen, non-spherical nebulae,
many of which reveal axi- and point-symmetries. For example,
the higher mass-loss super-wind that marks the end of the AGB phase
may be highly asymmetric, and in the Interacting Stellar Wind
model could result in a non-spherical nebula.
The young nebula of He 2-138
is undoubtedly complex (e.g. Sahai {\&} Trauger 1998) and
suggestive of strongly collimated post-AGB mass-loss.
The morphology of the low and high ion UV lines presented here
also points to an asymmetry in the subsequent central star wind.
The cause of this asymmetry remains uncertain. The role of
subsurface convection and a sustained AGB magnetic dynamo has
been studied previously in the context of bipolar PN shapes
(e.g. Blackmann et al. 2001; Nordhaus et al. 2007). The combination
of rotation and a long-term large-scale magnetic field might
then lead to an equatorially enhanced super-wind.
Indeed our notion that the central star in He 2-138 may be
a rapid rotator would be consistent with the simulations of
e.g. Dwarkadas (2004), who reported that fast rotation of
the central star can lead to non-spherical mass-loss. In the
early stages of the nebular evolution the aspherical mass-loss
could then result in a bipolar wind-blown bubble.

An alternative scenario is that binarity is the cause of the
non-spherical planetary nebulae. Photometric and spectroscopic
radial velocity surveys have suggested that at least $\sim$ 15{\%}
of PN central stars are in short period (less than 3 days)
close binaries (e.g. Bond 2000; De Marco et al. 2004).
Unambiguous spectroscopic confirmation of central star binaries
has been difficult to achieve and in many cases the fluctuations
of deep-seated spectral lines appear to be erratic and not
cyclic on a clear period (e.g. De Marco et al. 2008).
In our study of He 2-138 the photospheric He{\sc i} and
metal lines are also variable, but they do not exhibit the
tell-tale sinusoidal pattern in radial velocity shifts. We argue
here instead that at least the $\sim$ hourly temporal changes evident in 
the photospheric
lines of He 2-138 are more reliably interpreted as
blue-to-red migrating features in the absorption profiles, which
may provide tentative evidence for photospheric velocity fields
and wave-leakage
due to prograde non-radial pulsations. We would not assign
the changes seen in e.g. He{\sc i} $\lambda$4026,
He{\sc i} $\lambda$4471, and C{\sc iv} $\lambda$5801 as
due to the effects of a variable central star {\it wind}.
In He 2-138 the outflow is only diagnosed in our data via
the P~Cygni-like profiles of He{\sc i} $\lambda$5876 and the
FUV and UV resonance lines.

Unravelling the occurrence of central star binarity and
photospheric structures (pulsational or magnetic) demands
intensive time-series over several days. These requisites are
observationally challenging to secure, particularly as
multi-site observations may also been needed in order to
provide sufficiently continuous monitoring. However, we have
demonstrated in this study of He 2-138 that it may be
rewarding to conduct an extensive comparative study of the
morphologies of the UV fast wind lines of young PN central stars, with
the goal of investigating in time-averaged (archival) data the
evidence for non-spherical outflows. These signatures may be
relics of the structure of the preceding post-AGB super-wind phase.

\section*{Acknowledgments}
We are grateful for the support of staff at the European
Southern Observatory, La Silla, Chile. Thanks also to
Adam Burnley for his support of the ESO observing run.
Finally, we would like to thank John Hillier for making the
CMFGEN code available to us.

\bsp

\label{lastpage}


\begin{thebibliography}{99}

\bibitem{}Bond, H. E. 2000, in ASP Conf. Ser. 199: Asym-
metrical Planetary Nebulae II: From Origins to
Microstructures, 115
\bibitem{}Afsar, M., Bond, H. E. 2005, Memorie della So-
cieta Astronomica Italiana, 76, 608
\bibitem{}Blackman, E. G., Frank, A., Markiel, J. A.,
Thomas, J. H.,Van Horn, H. M. 2001, Nature, 409, 485
\bibitem{}Cranmer, S. 1996, Ph.D. Thesis, University of Delaware
\bibitem{}De Marco, O., Bond, H.E., Harmer, D., Fleming, A.J. 2004,
ApJ, 602, 93
\bibitem{}De Marco, O., Hillwig, T. C., Smith, A. J.
2008, AJ, 136, 323
\bibitem{}Dwarkadas, V.V., 2004, in Asymmetrical Planetary Nebulae III: 
Winds, Structure and the Thunderbird, Proceedings of the conference held 
28 July - 1 August 2003 at Mt. Rainer, Washington, USA. eds. M. 
Meixner, J. H. Kastner, B. Balick and N. Soker. ASP Conference 
Proceedings, Vol. 313. San Francisco: Astronomical Society of the Pacific, 
2004., p.430
\bibitem{}Fullerton, A.W., Gies, D.R., Bolton, C.T, 1996, 
ApJS, 103, 475 \bibitem{}Hillier, D.J., Miller, D., 1998, ApJ, 496, 407
\bibitem{}Hubeny, I., Lanz, T. 1995, ApJ, 439, 905
\bibitem{}Hutton, R.G., Mendez, R.G. 1993, A{\&}A, 267, 8
\bibitem{}Kastner, J.H., Montez, R., Jr., Balick, B.,
De Marco, O. 2008, ApJ, 672, 957
\bibitem{}Kudrtizki, R.P., Urbaneja, M.A., Puls, J. 2006, IAU Symp. 234,
'Planetary Nebulae in our Galaxy and
Beyond', eds. M.J. Barlow {\&} R.H. Mendez, CUP, 119.
\bibitem{}Lamers, H.J.G.L.M., Cerrutti-Sola, M., Perinotto, M. 1987,
ApJ, 314, 726
\bibitem{}Massa, D., Fullerton, A.W., Sonneborn, G., Hutchings,
J.B. 2003, ApJ, 586, 996
\bibitem{}Mendez, R.H., Forte, J.C., Lopez, R.H. 1986,
Rev. Mex. Astron. Astrofis., Vol. 13, No. 2, 19 
\bibitem{}Mendez et al. 1988, A{\&}A,190, 113
\bibitem{}Nordhaus, J., Blackman, E. G., {\&} Frank, A. 2007,
MNRAS, 376, 599
\bibitem{}Perinotto, M. 1987, IAU Symp. 131, 293
\bibitem{}Prinja, R.K., Hodges, S.E., Massa, D.L.,
Fullerton, A.W, Burnley, A.W. 2007, MNRAS, 382, 299
\bibitem {}Roberts, D.H., Leh{\'a}r, J., Dreher, J.W. 1987, AJ,
93, 968
\bibitem{}Sahai, R., Trauger, J.T.  1998, AJ, 116, 1357
\bibitem{}Schneider, S.E., Terzian, Y., Purgathofer, A.,
Perinotto, M. 1983, ApJS, 52, 399
\bibitem{}Townsend, R.H.D. 2000, MNRAS, 318, 1
\bibitem{}Urbaneja , M.A., Kudritzki, R.-P.,
Puls, J., 2008, in Clumping in hot-star winds : proceedings of an 
international workshop held in Potsdam, Germany, 18. - 22. June 2007. 
Hamann, Wolf-Rainer (ed.) ; Feldmeier, Achim (ed.) ; Oskinova, Lidia M. 
(ed.). ISBN 978-3-940793-33-1., p.67
\bibitem{}Zhang, C.Y. 1995, ApJS, 98, 659 
\bibitem{}Zickgraf, F.-J., Wolf, B., Stahl, O. 1985, A{\&}A,
143, 421

\end{thebibliography}
\end{document}